\documentclass[aps,prl,twocolumn,superscriptaddress,showpacs,showkeys,amsmath,amssymb]{revtex4}
\usepackage{graphicx}
\usepackage{dcolumn}
\usepackage{bm}
\usepackage{color}
\usepackage[thinspace,thinqspace,amssymb,pstricks]{SIunits}
\usepackage{longtable}
\usepackage[letterpaper,hypertex]{hyperref}
\hypersetup{
    pdftitle={Frontiers in Physics},
    pdfauthor={Haifeng Ma, Thomas Greber},
    pdfsubject={State of endohedral Ar in C$_{60}$.}, pdfkeywords={Endofullerenes, Photoemission, C$_{60}$}}

\begin{document}

\title{Corrugated single layer templates for molecules: From $h$-BN Nanomesh to Graphene based Quantum dot arrays}
\date{\today}

\author{Haifeng Ma}
\affiliation{Physik-Institut, Universit\"{a}t Z\"{u}rich, Winterthurerstrasse 190, CH-8057 Z\"{u}rich, Switzerland}
\author{Mario Thomann}
\affiliation{Physik-Institut, Universit\"{a}t Z\"{u}rich, Winterthurerstrasse 190, CH-8057 Z\"{u}rich, Switzerland}
\author{Jeanette Schmidlin}
\affiliation{Physik-Institut, Universit\"{a}t Z\"{u}rich, Winterthurerstrasse 190, CH-8057 Z\"{u}rich, Switzerland}
\author{Silvan Roth}
\affiliation{Physik-Institut, Universit\"{a}t Z\"{u}rich, Winterthurerstrasse 190, CH-8057 Z\"{u}rich, Switzerland}
\author{Martin Morscher}
\affiliation{Physik-Institut, Universit\"{a}t Z\"{u}rich, Winterthurerstrasse 190, CH-8057 Z\"{u}rich, Switzerland}
\author{Thomas Greber}
\email{greber@physik.uzh.ch}
\affiliation{Physik-Institut, Universit\"{a}t Z\"{u}rich, Winterthurerstrasse 190, CH-8057 Z\"{u}rich, Switzerland}

\begin{abstract}
Functional nano-templates enable self-assembly of otherwise impossible arrangements of molecules. A particular class of such templates is that of $sp^2$ hybridized single layers of hexagonal boron nitride or carbon (graphene) on metal supports.
 If the substrate and the single layer have a lattice mismatch, superstructures are formed. 
On substrates like rhodium or ruthenium these superstructures have unit cells with $\sim$3 nm lattice constant.
They are corrugated and contain sub-units, which behave like traps for molecules or quantum dots, which are small enough to become operational at room temperature. 
For graphene on Rh(111) we emphasize a new structural element of small extra hills within the corrugation landscape.
For the case of molecules like water it is shown that new phases assemble on such templates, and that they can be used as "nano-laboratories" where many individual processes are studied in parallel. 
Furthermore, it is shown that the $h$-BN/Rh(111) nanomesh displays a strong scanning tunneling microscopy induced luminescence contrast within the 3 nm unit cell which is a way to address trapped molecules and/or quantum dots.
\end{abstract} 

\pacs{}

\keywords{hexagonal boron nitride, graphene, nano-template, quantum dot, nano-ice, nanomesh, electroluminescence}

\maketitle

\section{Introduction}

Graphene resounds throughout the land \cite{gei07}. It is a single layer of $sp^2$ hybridized carbon, has remarkable chemical stability and physical properties like that of a conductor with high charge carrier mobility. Her polar sister, hexagonal boron nitride, has similar chemical stability, though is an insulator.

Single layer hexagonal boron nitride ($h$-BN) and graphene ($g$) also have great potential as templates for molecular self-assembly.
The layers are grown and supported on transition metal surfaces \cite{osh97,gre103}.
Here we focus on \emph{corrugated} single layers.
The corrugation is a vertical deformation of the surface that can be described as a static distortion wave. 
The physical origin of these distortions are the mismatch and the concomitant epitaxial stress between the overlayer and the substrate, where the
anisotropic bonding or lock in energy imposes this kind of dislocations. The much softer out of plane modulus of the $sp^2$ layers causes a large vertical distortion compared to the in-plane straining.
The wavelengths, or superlattice constants, of these static distortion waves can be calculated from the lattice mismatch between the overlayer and the substrate. 
For rhodium and ruthenium it is in the order of 3 nm and the corrugation, or peak to peak amplitude (between 0.05 and 0.15 nm) are the essential features and determine the template function.
It has been shown that the corrugation imposes \emph{lateral} electric fields, which can guide charged or polarizable media \cite{dil08,bru09}. 
This property leads to 3 bond hierarchy levels. The $\sigma$-bonds, in the order of 10 eV provide chemical stability and robustness, the $\pi$-bonds, in the order of 1 eV, the adsorption energies, and the $\alpha$-bonds(named after the label $\alpha$ for the polarizability $\alpha$), in the order of 100 meV, are responsible for the lateral trapping of molecules \cite{gre101}.

This article covers the basic ingredients of the geometric structure of lattice mismatched $sp^2$ layers on transition metals, their potential as "nano-laboratories", where an example of the behavior of nano-ice clusters as a function of temperature is given. Finally, the potential of such superstructures as quantum dot arrays is outlined. It is shown that such quantum dots can be addressed by electroluminescence, where the yield varies one order of magnitude within the 3 nm unit cell of $h$-BN/Rh(111).

\section{Geometric structure}

When the lattice mismatch $M$ of an overlayer with the substrate exceeds a
critical value, superstructures with large lattice constants are formed. 
For parallel epitaxy we write:
\begin{equation}
M=\frac{a_{ovl}-a_{sub}}{a_{sub}}
\label{E1}
\end{equation}
where $a_{ovl}$ and $a_{sub}$ are the overlayer  and the substrate lattice constants, respectively.
In this notation positive (negative) $M$ indicate compressive (tensile) stress in the overlayer, and vice versa.
For most transition metal substrates with $h$-BN or $g$ overlayers the mismatch is negative, i.e. tensile stress in the overlayer prevails.

\begin{figure}
\includegraphics[width=0.8\columnwidth]{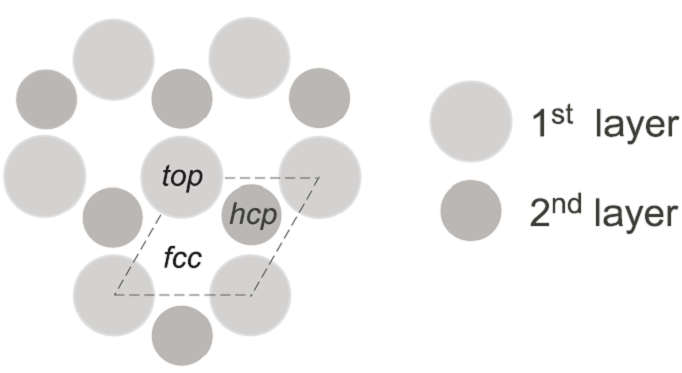} 
\caption{Scheme for the classification of the adsorption sites on a hexagonal closed packed surface: $top$, $fcc$, and $hcp$. The dashed line shows the (1$\times$1) unit cell. After Ref. \cite{auw99}}
\label{Ftopfcchcp}
\end{figure}

If the lattice of the overlayer and the substrate are rigid and parallel, the superstructure lattice constant gets $a_{ovl} / |M|$, where $a_{ovl}$ is the 1$\times$1 lattice constant of  $h$-BN or graphene ($\sim$ 0.25 nm).
Besides the mismatch, the balance between the lock in energy and the strain energy is decisive for the resulting morphology of the systems. 
Lock in energy has to be paid when the over-layer atoms are moved parallel to the surface, away from the lowest energy sites.
For systems with small lock in energy compared to the strain energy,
we expect flat floating layers, reminiscent to incommensurate moir\'e
patterns without a lock in to a high symmetry direction of the substrate.
Such examples of moir\'e's are e.g. $h$-BN/Pd(111)\cite{mor06,gre09} or $g$/Ir(111) \cite{ndi06}.
However, if the unit cells of the superstructure contain regions with distinct electronic structure, as it is e.g. the case for $h$-BN/Rh(111) \cite{cor04,ber07} or $g$/Ru(0001) \cite{bru09,zha10}, it is appropriate to use a term distinct from moir\'e, like it is 'nanomesh'. 
Preobrajenski et al. were the first who also used the term nanomesh for $g$-systems where two distinct carbon core levels have been found \cite{pre081}.
The energy difference in the core level binding energies was assigned to the corrugation, i.e. different 'elevations' or distances of the $sp^2$ layers from the substrate.
For a superstructure of a honeycomb lattice of $sp^2$ hybridized layers on a hexagonal closed packed substrate as Rh(111) or Ru(0001), it is convenient to describe different locations along the notation used by Auw\"arter and Grad et al.  \cite{auw99,gra03}. The honeycomb lattice is made of a base with two atoms  (B,N) or (C$_A$,C$_B$). These atoms may sit on $top$, on $fcc$ or on $hcp$ sites within the substrate unit cell (see Figure \ref{Ftopfcchcp}). The $top$ site is occupied if an atom of the honeycomb sits on top of a substrate atom, below the $fcc$ hollow site no atom is found in the second substrate layer, while there is one for the $hcp$ hollow site. 
We note that for single domain $h$-BN structures only three of the six combinations of the honeycomb base like ($top$,$fcc$), ($hcp$,$top$), ($fcc$,$hcp$) or  ($fcc$,$top$), ($top$,$hcp$), ($hcp$,$fcc$) occur.
For graphene, where the two carbon atoms are distinct by their coordination to the substrate only, no such domains are expected.
When the honeycomb is mismatched, i.e. does not have the lattice constant of the substrate, the assignments (B,N)=($fcc$,$top$),($hcp$,$fcc$) etc. are only approximately valid and we write e.g. (C$_A$,C$_B$)$\sim$($top$, $hcp$).
With this scheme in mind we may understand the complementarity of mismatched $h$-BN and $g$ 'nanomeshes', where it is found that about one third of the super cell (B,N)$\sim$($hcp$,$fcc$) and (C$_A$,C$_{B}$)$\sim$($hcp$,$fcc$) do not bind to the substrate and consequently belong to the elevated regions. For $h$-BN there is also no bonding for boron on $top$ sites i.e. (B,N)$\sim$($top$,$hcp$), and consequently two thirds of the mismatched $h$-BN layers are elevated and form the connected 'wire' network. 
Graphene is complementary i.e. the (C$_A$,C$_{B}$)$\sim$($top$,$hcp$) and the (C$_A$,C$_{B}$)$\sim$($fcc$,$top$) sites bind to the substrate and form a  connected hexagonal 'valley' network, with graphene in close contact to the substrate \cite{bru09}. It has to be emphasized that the substrate breaks the symmetry between the sublattice made of C$_A$ and C$_B$ atoms, respectively, and disables the formation of Dirac cones, which are the attribute of freestanding graphene. For $h$-BN this symmetry breaking is intrinsic, since B and N are different and induce polarity with electron transfer to nitrogen. 

\begin{figure}
\includegraphics[width=1\columnwidth]{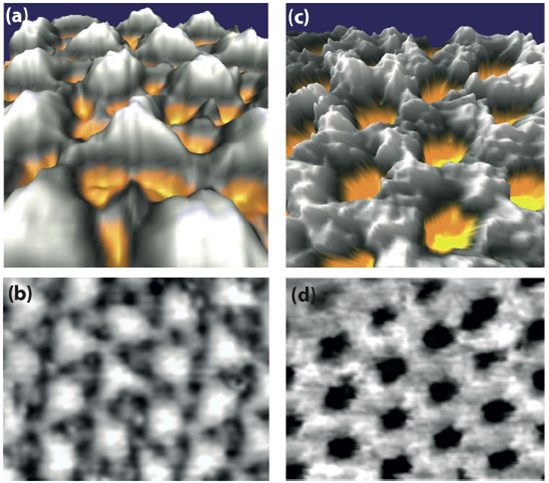} 
\caption{Scanning tunneling microscopy data of $sp^2$ single layers on Rh(111). a) Relief view of $g$/Rh(111). Note the hills and the valleys, and the small extra hills at the ($hcp$,$fcc$) sites. b) corresponding STM picture ($I_t= 0.8$ nA, $U_t= -0.8$ V , 11$\times$14 nm$^2$).
c) Relief view of $h$-BN/Rh(111) nanomesh. The elevated regions form the so called wires of the nanomesh. d) corresponding STM picture ($I_t= 1$ nA, $U_t= 1$ V, 11$\times$14 nm$^2$).}
\label{Fstruct}
\end{figure}

Figure \ref{Fstruct} shows room temperature scanning tunneling microscopy (STM) pictures of $g$/Rh(111) and $h$-BN/Rh(111). 
The relief views in a) and c) are extracted by the WSxM Scanning Probe Microscopy software \cite{hor07} from the scanning tunneling microscopy data in b) $h$-BN/Rh(111) and d) $g$/Rh(111), respectively. Clearly, the inverted topographies of the two layer systems can be seen.
The $h$-BN/Rh(111) nanomesh has a 12$\times$12 periodicity where 13 BN units sit on 12 Rh substrate units, which corresponds to a 3.2 nm superlattice constant \cite{cor04}.
The labels for the two topographic elements are 'holes', 'pores', 'cavities' or 'cells' for the (B, N)=($fcc$,$top$) regions with close binding and 'wires' for the  (B, N)=($hcp$,$fcc$) \& ($top$,$hcp$) regions, which are elevated by about 0.1 nm.
The $g$/Rh(111) 'nanomesh' has a periodicity of about 11$\times$11, where 12 $g$ units sit on 11 Rh substrate units, which corresponds to a 2.9 nm superlattice constant \cite{mul09}. The slightly smaller unit cell is due to the smaller lattice constant of graphene compared to that of hexagonal boron nitride.
The labels for the two topographic elements are 'mounds', 'hills' or 'ripples' for the ($C_A$, $C_B$)=($hcp$,$fcc$) protrusions with loose binding and 'valleys' for the  ($C_A$, $C_B$)=($fcc$,$top$)  \& ($top$,$hcp$) regions, which are about 0.1 nm closer to the substrate.
For the case of $g$/Rh(111) we would like to mention a difference, compared to the related $g$/Ru(0001) system. It can be seen that the strongest bonding does not coincide with the ($fcc$,$top$) or ($top$,$hcp$) sites but 3 small extra dips in the valleys where carbon atoms are closer to bridge sites are binding strongest to the substrate \cite{iannuzzi}.
This, compared to $g$/Ru(0001), new structural element might impose extra effects in the template function and should be further explored.

As it was inferred by photoemission from adsorbed xenon for $h$-BN/Rh(111) \cite{dil08} and $g$/Ru(0001) \cite{bru09}, in both systems 'high' or 'elevated' regions have a high local work function while low regions have a lower local work function \cite{bru09}. 
These physically and electronically corrugated landscapes form templates for the self-assembly of molecular arrays as it was shown for $h$-BN nanomesh \cite{cor04,ber07,mah10}, $g$/Rh(111) \cite{pol10} or $g$/Ru(0001) \cite{mao09}.
Also it was demonstrated that these substrates may be used for the growth of metal nano-clusters \cite{zha08,pan09}.

\section{Nano-laboratories: Water on the $h$-BN nanomesh}

\begin{figure*}
\includegraphics[width=2\columnwidth]{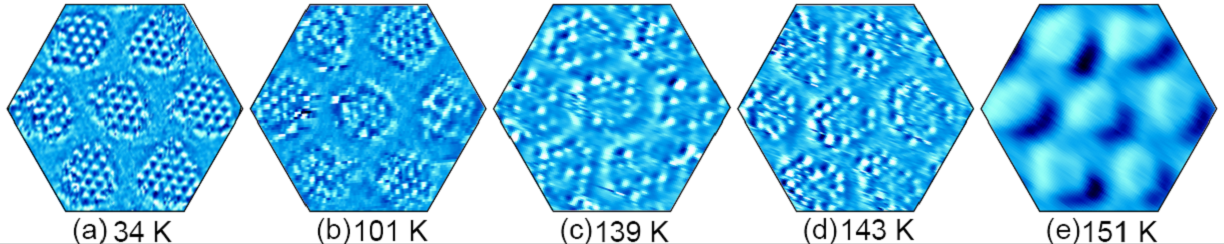} 
\caption{\label{F1} Behavior of nano-ice clusters in the temperature range between 34 K and 151 K as recorded by variable temperature scanning tunneling microscopy (VT-STM).
The 5 hexagonal frames with a side length of 4.2 nm show the constant current feed-back signal, after drift correction. (a) Nano-ice clusters at 34 K. $V_t= -0.05$ V, $I_t= 100$ pA. (b) At 101 K water molecules at the rims of the nano-ice cluster become mobile, the core remains frozen. $V_t= -0.05$ V, and $I_t= 100$ pA. (c) Low coordinated water induced protrusions at the cluster boundary are observed at 139 K, the core features weaken. $V_t=  -0.01$ V, and $I_t= 50$ pA. (d) as (c), the rims show higher contrast than the core molecules which still show crystallinity at 143 K. $V_t= -0.01$ V, and $I_t= 50$ pA. (e) Empty nanomesh after water desorbed from the surface. $V_t= -0.05$ V, and $I_t= 40$ pA.}
\label{Fice}
\end{figure*}

Here we want to highlight the opportunity to use a template like the $h$-BN nanomesh as a nano-laboratory, where processes may be studied in a parallel fashion, i.e. in an ensemble, at the same time, under the same temperature and pressure conditions. For scanning probes this also comprises the added value that the data are recorded with the same tip. These features increase the data flux from the experiment by orders of magnitude. In particular we expect that e.g. the diversification in growth processes may be studied. 
This nano-laboratory assay allows, to study equilibrium as well as non equilibrium processes. If we want to lend a picture from biology the $h$-BN nanomesh can be considered as a "cell culture", though at a 3 orders of magnitude smaller length scale.  
In order to illustrate the "nano-laboratory" we show data on the temperature evolution of water clusters in the $h$-BN nanomesh.
Recently it was shown that water self-assembles in the 'holes' of  $h$-BN nanomesh in forming ice nanoclusters containing about 40 water molecules. The clusters consist in a bilayer of ice, reminiscent to the basal plane of hexagonal ice, and display proton disorder that was accessed with tunneling barrier height measurements \cite{mah10}.
In Figure \ref{Fice} the development of the ice clusters as a function of temperature is shown, for 5 different temperatures, About 7 nanomesh "cells" with a diameter of 2 nm contain each one ice cluster. It has to be mentioned that the results for different temperatures do not show the same cells, because the thermal drift in the present set up during the warm up from 34 to 151 K does not yet allow to track individual cells. 
The superstructure periodicity does, however, allow an almost perfect drift correction at a given temperature.
It can be clearly seen that the rims of the ice clusters have a distinct behavior with respect to the bulk.
Figure \ref{Fice} a) shows the ordered ice clusters made by about 40 water molecules at 34 K. Every second water molecule shows up as a protrusion inside the $h$-BN nanomesh.
At 101 K the edge of the clusters start to display disorder (Figure \ref{Fice} b).
 Further increase of the temperature induces an increase of the cluster height (Figure  \ref{Fice} c) and d)) and at 151 K the clusters sublimated (Figure \ref{Fice} e) and bare nanomesh is observed.

\section{Addressing Room temperature Quantum dots}

A quantum dot is a small physical object that is confined in 3 dimensions, where nuclei and atoms are the most prominent examples.  The localization to a 'point' implies the absence of dispersion of the electronic states. The size of a quantum dot determines the energy level spacing. If this spacing $\Delta E$ is compared to $k_B T$ we get a measure for the temperature below which we expect the 'dots' to behave like quantum objects, or above which the occupancy of different levels fluctuates.
The Rydberg energy (13.6 eV), which is the scale for the electronic level spacing in a Coulomb potential of a proton is proportional to ${a_o}^{-1}$, where $a_0$=0.05 nm is the Bohr radius. 
If we are interested in quantum dots that are operational at room temperature ($k_B T$=25 meV), this limits the size of quantum dots to below 500 $a_o$. Though, for practical purposes the level spacing should be at least one order of magnitude larger than $k_B T$ and thus room temperature quantum dots should be smaller than 5 nm.

The $sp^2$ templates discussed in this paper do match this condition. Indeed quantum dot behavior was found for graphene on ruthenium \cite{zha10}. For $g$/Ru(0001) photoemission showed one set of of $sp^2$ valence bands, while $h$-BN/Ru(0001) and $h$-BN/Rh(111) do show two $sp^2$ valence band structures split by about 1 eV \cite{gor07}. The two band structure systems were assigned to the two regions within the super cells, where the corrugation of the $h$-BN imposed, a mainly electrostatically driven splitting \cite{las07,ber07}. As expected this splitting is also observed with high resolution B 1s, C 1s and N 1s core level spectroscopies \cite{pre081}, and photoemissions from adsorbed Xe \cite{dil08,bru09}. 
Interestingly the valence band splitting was not observed for $g$/Ru(0001) \cite{bru09}. The obvious difference between graphene and $h$-BN is that graphene on ruthenium has a Fermi surface, while $h$-BN on Ru(0001) \cite{bru09} or on Rh(111) \cite{gre09} has not, may not explain this with a screening argument, since the C 1s core level is still split on $g$/Ru(0001) \cite{pre081}. The seeming paradox can be resolved, when we assign to the hills in the $g$/Ru(0001) superstructure an isolated, molecule like behavior, without dispersion, which qualifies them as quantum dots arranged on a hexagonal array with 3 nm spacing. 
For the case of the $h$-BN, the holes might also be identified as quantum dots, however, angular resolved photoemission shows dispersion of the $h$-BN valence bands, also for the bands assigned to the holes that are separated by the superlattice constant of 3 nm \cite{bru09}. 
A difference between the hills of $g$/Ru(0001) and the holes of $h$-BN/Ru(0001) is the fact that $h$-BN holes are in close contact to the substrate, while graphene hills are decoupled. This imposes a lateral vacuum tunneling barrier for electrons on the hills, while this barrier is much lower for the case of the $h$-BN holes that are in close proximity to the metal substrate.

If the $sp^2$ templates are decorated with molecules (or clusters), the quantum dots change and the coupling between them will be affected. It will be interesting to further explore this coupling and to try to control it.
Electroluminescence could serve as a tool to access such information.
With scanning tunneling microscopy induced electroluminescence, light emission can be probed as a function of the tunneling site with sub-wavelength resolution.
As we show here it is one possible road to access single unit cells and is considered to be a realization of a nano-device, where it comes to the transport of information localized at the nanometer scale to the macroscopic millimeter scale.  

\begin{figure}
\includegraphics[width=0.9\columnwidth]{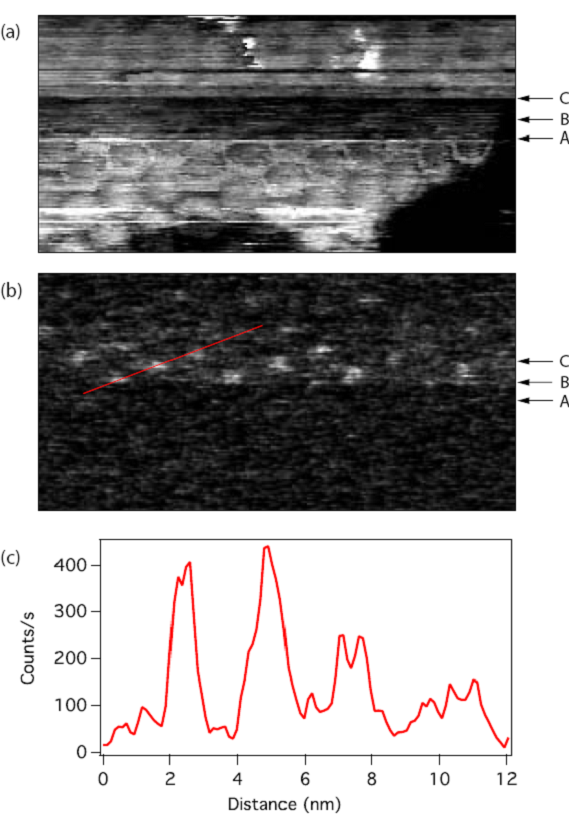} 
\caption{Room temperature scanning tunneling microscopy and photon emission scanning tunneling microscopy from $h$-BN nanomesh with a gold coated tungsten tip. (30$\times$15 nm, $I_t= 2.6$ nA, $V_t= -2.5$ V, scan time 110 s with 128 horizontal scanning lines from bottom to top). The labels $A, B, C$ indicate tip changes (for details see text).
a) Topography, note the 3 nm periodicity of $h$-BN nanomesh, and the tip changes $A$, $B$. b) Light map, i.e. simultaneously to a) recorded photoncurrent.  For a certain line series the luminescence is particularly high, and the periodicity of the $h$-BN nanomesh lights up. c) Cut across the light map, along the red line in b). The polychromatic light current is given in photons/s. }
\label{Flumi}
\end{figure}

Electroluminescence in scanning tunneling microscopy was pioneered by Gimzewski et al. \cite{gim88}, and is a way to record inelastic scattering in tunneling junctions, where e.g. molecular vibrations may be resolved, if the wavelength of the emitted photons is measured \cite{wuw06}.
Figure \ref{Flumi} shows the correlation between topography and light emission from $h$-BN nanomesh. Light was collected by a lens system connecting the tunneling junction with a cooled red sensitive Hamamatsu R5929 photomultiplier tube operating in the wavelength window between 300 and 850 nm. The tunneling voltage was set to -2.5 V (tunneling electrons from the substrate to the tip) and tungsten tips with a 80 nm gold coating were used.
For isotropic emission into the $2\pi$ half space above a surface, the detection probability of a 2 eV photon is 0.4 \%, and the dark count rate was about 25 counts/s. The photon map in Figure \ref{Flumi} b) and the corresponding cut in Figure \ref{Flumi} c) show strong electroluminescence from the nanomesh 'holes', which is more than one order of magnitude larger than that from the wires. Though, the average quantum efficiency (detected photons per scan line) varies in the shown data by one order of magnitude. This variation must be ascribed to changes in the gold tip where plasmon excitations/deexcitations cause photon emission. 
The arrows on the right of Figure \ref{Flumi}a) and b) indicate three distinct tip changes $A, B, C$. 
It can be seen that the topography image before change $A$ is inverted after change $C$. The enhancement of the quantum efficiency at change $B$ does neither coincide with $A$ nor $C$.
However, the results suggest that the tunneling junction with the tip on top of a hole of the $h$-BN nanomesh imposes more inelastic scattering events on the tip. The high electron affinity and the concave form of the holes lets them act like a resonator cavity, where the electrons are focused on the tip, and where the probability for a plasmon excitation increases. It has to be mentioned that the inverse situation is observed for graphene on Ru, where the poor electron affinity and the convexity of the hills defocus electrons in a tunneling junction with the tip on top of the hill \cite{zha10}.

The data shown in Figure \ref{Flumi} indicate that scanning tunneling microscopy induced luminescence can be used for the identification of sites within the 12$\times$12 super cell of $h$-BN/Rh(111) with sub-nanometer resolution. Also the experiments indicate that the control of the tip parameters is crucial for a successful application of this effect.

\section{Summary}
In summary  it is recalled  that lattice mismatched $sp^2$ hybridized single layers of $h$-BN and graphene may be used as templates for the self-assembly of molecular structures.
For $g$/Rh(111) a new structural element, extra "hills" in the valleys, are emphasized.
In a second section it is shown that $h$-BN/Rh(111) can be used as a "nano-laboratory", where molecular processes in individual nanomesh cells may be studied. 
Finally it is outlined that these superstructures have the features of quantum dots, which are small enough to become operational at room temperature. As an example on how such objects can be addressed the luminescence as induced by a scanning tunneling microscope is demonstrated to have a resolution better than one nanometer.

\section{Acknowledgements}
Financial support by the Swiss National Science Foundation is gratefully acknowledged.

\end{document}